\documentstyle[12pt]{article}
\def\beq{\begin{equation}}
\def\eeq{\end{equation}}
\def\beqn{\begin{eqnarray}}
\def\eeqn{\end{eqnarray}}
\def\lapproxeq{\lower .7ex\hbox{$\;\stackrel{\textstyle <}{\sim}\;$}}
\def\gapproxeq{\lower .7ex\hbox{$\;\stackrel{\textstyle >}{\sim}\;$}}

\def\as{\alpha_s}

\def\lmsb{\Lambda_{\overline{\rm MS}}}

\def\GeV{{\rm GeV}}
\def\MeV{{\rm MeV}}
\def\TeV{{\rm TeV}}
\def\qq{q \bar{q}}

\def\pp{p\bar p}

\def\half{{\textstyle{1\over 2}}}
\def\quarter{{\textstyle{1\over 4}}}

\def\d{{\rm d}}
\def\sigmahat{\hat{\sigma}}

\def\tf{{\tilde f}}
\begin{document}
\begin{titlepage}
\vspace*{-1cm}
\begin{flushright}
DTP/94/04 \\
January 1994\\
\end{flushright}
\vskip 1.cm
\begin{center}
{\Large\bf
Production of Jet Pairs at Large Relative Rapidity
 in Hadron-Hadron Collisions as a
Probe of the Perturbative Pomeron}
\vskip 1.5cm
{\large  W.J. Stirling}
\vskip .5cm
{\it Departments of Mathematical Sciences and Physics, University of Durham \\
Durham DH1 3LE, England }\\
\vskip 1.5cm
\end{center}
\begin{abstract}
The production of  jet pairs with small transverse
momentum and large relative rapidity in high energy hadron-hadron
collisions is studied. The rise of the parton-level cross section with
increasing
rapidity gap is a fundamental prediction of the BFKL \lq perturbative
pomeron' equation of Quantum Chromodynamics. However, at fixed collider
energy  it is difficult
to disentangle this  effect from variations in the cross section due to
the parton distributions. It is proposed  to study instead the
 distribution in the azimuthal angle difference of the jets
 as a function
of the rapidity gap.  The flattening of
this distribution with increasing dijet rapidity
gap is shown to be a characteristic feature of the BFKL behaviour.
Predictions for the Fermilab $\pp$ collider
are presented.
\end{abstract}
\vfill
\end{titlepage}
\newpage
%
%
\section{Introduction}

There is currently much interest in the
QCD \lq perturbative pomeron'.  This is the phenomenon,  obtained by
resumming a certain type of soft gluon emission to all orders in the
leading logarithm approximation  using the
 Balitsky-Kuraev-Fadin-Lipatov (BFKL) equation \cite{BFKL}, which is supposed
to produce a   sharp rise  at small $x$ in deep inelastic  structure
functions \cite{JAN}. Recent measurements  at HERA \cite{HERA}
may well show the  first evidence for this behaviour,
although it is premature to draw any definitive conclusions.

One of the difficulties with extracting information on the perturbative
pomeron from structure function measurements alone is that both
perturbative and non-perturbative effects are very likely intertwined in a
non-trivial way, and therefore obtaining information on the former
requires some sort of model-dependent subtraction of the latter.
For this reason, attempts have been made to find other quantities which
probe more directly the perturbative behaviour.
In the context of deep inelastic scattering, one can look for an
associated jet of longitudinal
momentum fraction $x' \gg x_{\rm Bj}$ and comparable
transverse momentum to the virtual photon momentum $Q$ \cite{DISJET}.
The BFKL behaviour is then reflected in the growth of the cross section
with $\log(x'/x_{\rm Bj})$. Alternatively, Mueller and Navelet have shown
\cite{MUENAV} that  in high-energy hadron-hadron
collisions one can utilize a two-jet inclusive cross section where the
jets have small transverse momentum and a large relative rapidity,
$\Delta y$. The rise in the cross section with increasing $\Delta y$ is
then controlled by the perturbative pomeron.

It is this latter process that we  investigate here. We  first
define the cross section introduced in Ref.~\cite{MUENAV}, and show why
it is difficult  to measure under present experimental conditions. We
 then consider an alternative but closely-related quantity, the
azimuthal angle correlation between jets at large relative rapidity,
 which we  argue is more promising from an experimental point of view.
Predictions are presented for the Fermilab $\pp$ collider at
$\sqrt{s} = 1.8\ \TeV$. Our results also apply to the corresponding
two-jet cross section in deep inelastic scattering as measured at HERA,
although
we shall not pursue this issue here.

Consider the inclusive two-jet cross section in $\pp$ (or $pp$)
collisions at energy $\sqrt{s}$, where
each jet has a minimum transverse momentum $M \ll \sqrt{s}$, and the
jets are produced with equal and opposite large rapidity
$\pm \half\Delta$.\footnote{The restriction to {\it equal} and opposite
rapidities is
not crucial and is made here only to simplify the discussion.}
Adapting the results of  Ref.~\cite{MUENAV}, the cross section for this can be
written in the form
\beq
 \left. {\d \sigma \over \d y_1 \d y_2 }\right\vert_{y_1 = -y_2 = \half\Delta}
\; \simeq\;
x_1 G(x_1, M^2)\; x_2 G(x_2, M^2)\
\sigmahat(\as(M^2), M^2, \Delta) \;
\label{basic}
\eeq
where
\beqn
G(x, \mu^2) & = &  g(x,\mu^2)\;  +\;
 {4 \over 9}\; \sum_q (q(x,\mu^2) + \bar{q}(x, \mu^2) ) \; , \nonumber \\
x_1\;  =\;  x_2 & = & {2M\over \sqrt{s}} \, \cosh(\half\Delta)\;
\simeq \;{M\over \sqrt{s}} \, e^{\Delta/2}\; ,  \nonumber \\
\sigmahat(\as, M^2, \Delta)
  &=& \left( {\as C_A \over \pi}\right)^2\;
 {\pi^3 \over 2 M^2} \; \left[ 1 + \sum_{n\geq 1} a_n \; (\as \Delta)^n +
\ldots \ \right] \; .
\label{defs}
\eeqn
Note the use of the \lq effective subprocess approximation',
appropriate here because of the dominance of small momentum transfer in the
subprocess $\hat{t}$  channel. The $\ldots$ in (\ref{defs}) refers to
corrections
outside
the leading logarithm approximation implicit in (\ref{basic}), i.e.
terms of order  $\as^n \Delta^{n-1}, \; \as^n \Delta^{n-2}, \; \ldots$
and power-correction terms suppressed by powers of $e^{-\Delta}$.

According to the BFKL \lq perturbative pomeron' analysis,
the subprocess cross section $M^2 \sigmahat$ is expected to rise at large
$\Delta$. The asymptotic behaviour, ignoring possible
 \lq parton saturation' effects \cite{MUENAV}, is predicted to be
\beq
M^2 \sigmahat \sim e^{\lambda \Delta}, \quad \mbox{as}\ \Delta \to
\infty \; ,
\label{lambda}
\eeq
with $\lambda$ a number of order 0.5.
This is the analogue of the expected $x^{-\lambda}$ growth of the $F_2$
structure function at small $x$ \cite{JAN}.

Now with $\sqrt{s} = 1.8\ \TeV$ and $M\sim 10\ \GeV$, the maximum
value of $\Delta $ is of order $10$, which might at first sight appear
sufficiently large to test the predicted BFKL behaviour.\footnote{In
fact, the predicted asymptotic behaviour appears to set in at relatively
modest values of $\Delta$, see Section 2 and Ref.~\cite{MUENAV}.}
The problem,  however, is the additional $\Delta $
dependence in the cross section induced by the $x$ dependence of the
parton distributions in (\ref{basic}). At large $\Delta$, the
behaviour of the cross section will be completely dominated
by the $x\to 1$ suppression of the parton distributions.
This can only be avoided by
increasing the energy  $\sqrt{s}$ of the collider as $\Delta$ is
increased, in such a way that $x_1$ and $x_2$ remain fixed, which is
 not an easy proposition in practice.

There are at least two ways around this difficulty. First, one could
argue that the parton distributions at medium-to-large $x$ are now
sufficiently well known that they can be factored out of the measured
cross section with sufficient accuracy. While this might be true for the
 quark component of $G(x,\mu^2)$, the gluon component is much less well
 known at large $x$. Even if it was, there is still
a problem with scale dependence --- the cross section in (\ref{basic})
is not known beyond leading logarithm accuracy, and so the choice of scale
in the parton distributions is somewhat arbitrary.

A second possibility is to use the distribution in the azimuthal angle
difference between the two jets as a signature for BFKL behaviour. When
$\Delta$ is small, the cross section is dominated by the lowest order
$2\to 2$ scattering subprocesses and the jets are back-to-back in the
transverse
plane.
As $\Delta$ increases, more and more soft gluons with transverse momentum
$\sim M$ are emitted in the rapidity interval between the two fast jets,
and the azimuthal correlation is gradually lost until, asymptotically,
there is no correlation at all. The change in the overall
weighting, due to variations in the parton distributions
 as $\Delta$ increases at fixed $\sqrt{s}$,
 has no effect on the shape of the azimuthal distribution.

 The study of the azimuthal correlation between the two jets at large rapidity
is the subject of the present analysis. We first of all derive the basic
formulae, which is an extension of the treatment in \cite{MUENAV}, and then
make numerical predictions for experimentally measurable quantities.
We shall demonstrate that
the weakening of the correlation should already be observable
at the Fermilab $\pp$ collider.

\section{BFKL formalism for dijet production}

We start by considering the subprocess cross section for inclusive  two-jet
(i.e. parton) production. We are interested in jets produced with equal
and opposite large longitudinal energy, $k_L \sim E$ where $E$ is the
parton beam energy in the parton-parton centre-of-mass,
 and small transverse momentum $k_T > M$ where
$M \ll E$ is a fixed cut-off. The rapidity  gap between the jets is then
$\Delta y \equiv \Delta \simeq 2 \log(E/M) \gg 1$.
 In practice,  $M$ will be  a number of order $10\ \GeV$
and the maximum value of $E$ is set by the kinematic limit $\sqrt{s}/2$.

At the Fermilab collider, a large fraction of such jet pairs
 are produced  in gluon-gluon
collisions, and so in what follows we focus on the subprocess $gg\to gg
+ X$. The quark initiated processes are reinstated afterwards using the
effective subprocess approximation.
 If we assume, to begin with, that we can ignore the running of the
strong coupling  and take $\as = \as(M^2)$ as fixed,
then the subprocess  cross section of Eq.~(\ref{basic}) can be written
as  \cite{MUENAV}
\beq
\sigmahat(\as,M^2,\Delta) = \left( {\as C_A\over\pi}\right)^2 \; {\pi^3\over
2M^2} \;
\int_{-\pi}^{\pi} \d \phi \; F(\phi, \Delta) \ .
\label{sig}
\eeq
Here we have introduced the variable $\phi = \pi -\phi_{jj}$  where
$\phi_{jj}$ is the azimuthal angle difference between the two jets. Thus
$\phi = 0$ corresponds to back-to-back jets in the transverse plane.
 In what follows
we will be particularly interested in the differential distribution
$\d\sigmahat / \d \phi$, which is proportional to  $F$.

The function $F$ in Eq.~(\ref{sig}) is simply the quantity
 $f(k_{T1},k_{T2} ,\Delta)$
 defined in Ref.~\cite{MUENAV},  integrated over $ M^2 <
k_{Ti}^2 < \infty$ at fixed $\phi$. The latter function satisfies
the BFKL equation
(see Appendix). The solution for $F$ is
\beq
F(\phi, \Delta) = {1\over 2\pi}\; \sum_{n = -\infty}^{+\infty} \;
e^{in\phi} \; C_n(t) \; , \quad t = {\as C_A\over \pi}\;\Delta \;  ,
\label{fourier}
\eeq
where
\beqn
C_n(t) &=& {1\over 2\pi}\; \int_{-\infty}^{+\infty} \; {\d z\over z^2
+ \quarter } \; e^{2 t \chi_n(z)}, \nonumber \\
  \chi_n(z) &=& {\rm Re}\,  [\, \psi(1) -\psi(\half(1+|n|)+iz)\,  ] \; ,
\label{coeffs}
\eeqn
and $\psi(x)$ is the logarithmic derivative of the gamma function.
Note that this result corresponds to a perturbative expansion in powers
of $\as\Delta \sim \as \log(E^2/M^2)$. It is these leading logarithms which
have been resummed by the BFKL equation. Substituting back in the
original expression for the cross section gives
\beq
 \left. {\d \sigma \over \d y_1 \d y_2 \d\phi}\right\vert_{y_1 = -y_2 =
\half\Delta}
\; \simeq\;
 x_1 G(x_1, M^2)\; x_2 G(x_2, M^2)\
\left( {\as C_A\over\pi}\right)^2 \; {\pi^3\over 2M^2} \;
F(\phi, \Delta) \ ,
\label{basica}
\eeq
with $x_1 = x_2 = 2M \cosh({\half\Delta})/\sqrt{s}$. Before performing a full
numerical calculation of this cross section, we discuss
 several important analytic results which
follow from Eqs.~(\ref{fourier},\ref{coeffs}).

\subsection{Comparison with exact lowest-order calculation\label{locomparison}}

Application of the BFKL formalism only makes sense
in a kinematic region where the
exact leading order ($2\to 2$) cross section is well-approximated by
the first term
in the perturbation series on the right-hand side of Eq.~(\ref{basica}).
The latter is obtained by setting $t=0$ in
Eq.~(\ref{coeffs}), which  gives
\beqn
C_n(0) & = & {1\over 2\pi}\; \int_{-\infty}^{+\infty} \; {\d z\over z^2
+ \quarter }  = 1 \nonumber \\
& \Rightarrow & F(\phi, \Delta) = \delta(\phi) \; .
\eeqn
When this is substituted in Eq.~(\ref{sig}), we obtain
\beq
M^2\; \left. \sigmahat\right\vert_{\rm LO}  =\half \pi \as^2 C_A^2
\label{lo}
 \ .
\eeq
Note  that in the $\Delta\to\infty$ limit the subprocess
cross section scales as $1/M^2$. In this section we rederive this result
starting
 from the exact ($2\to 2$) lowest order cross section, which will enable us
to assess how rapidly the asymptotic behaviour is approached.

The two-jet inclusive cross section in leading order is given by
\beq
{\d\sigma\over \d y_1 \d y_2 \d p_T^2} = {1\over 16\pi^2 s}\;
\sum_{a,b,c,d =q,g}\;x_1^{-1} f_a(x_1,\mu^2)\;x_2^{-1} f_b(x_2,\mu^2)\;
\overline{\sum}\;| {\cal M}(ab\to cd)|^2 \;,
\eeq
where $\overline{\sum}$ denotes the appropriate sums and averages over colours
and spins. To begin with, we restrict our attention to the $gg\to gg$
subprocess.
Setting $y_1  =-y_2 =\half\Delta$ gives
\beq
\left. {\d\sigma\over \d y_1 \d y_2 \d p_T^2}\right\vert_{y_1 = -y_2 =
\half\Delta} =
{ x g(x,\mu^2)\;x g(x,\mu^2) \over 256  \pi\, p_T^4 \,
 \cosh^4(\half\Delta) }\
\overline{\sum}\;| {\cal M}(gg\to gg)|^2 \;,
\eeq
where
\beq
x =  {2 p_T \over \sqrt{s}}\; \cosh(\half\Delta) \; ,
\eeq
and the subprocess matrix element is evaluated at
\beq
{\hat{t} \over \hat{s} } = - {1\over 2}\left( 1 - \tanh(\half\Delta)
 \right) \; , \qquad
{\hat{u} \over \hat{s} } = - {1\over 2} \left( 1 + \tanh(\half\Delta)
 \right)  ; .
\eeq
Integrating over the jet transverse momentum $p_T > M$ then gives
\beq
\left. {\d\sigma\over \d y_1 \d y_2 }\right\vert_{y_1 = -y_2 =
\half\Delta} =
 { \overline{\sum}\;| {\cal M}(gg\to gg)|^2  \over
  256  \pi \, \cosh^4(\half\Delta) } \
 \int_{M^2}^{p_T^2({\rm max})} \; {\d p_T^2 \over p_T^4} \;
 [ x g(x,\mu^2) ]^2\; ,
\label{twototwo}
\eeq
with $p_T^2({\rm max}) =\quarter s \cosh^{-2}(\half\Delta)$.
By extracting a factor of $M^{-2}$, and ignoring logarithmic scaling violations
in the parton distributions, the integral on the right-hand
side  becomes a function
of the dimensionless quantity $X = 2M\cosh(\half\Delta)/\sqrt{s}$.
\beq
\left. {\d\sigma\over \d y_1 \d y_2 }\right\vert_{y_1 = -y_2 =
\half\Delta} =
 { \overline{\sum}\;| {\cal M}(gg\to gg)|^2  \over
  256  \pi \, \cosh^4(\half\Delta) } \ {1\over M^2}\
 \int_1^{X^{-2}} \; {\d u^2\over u^4} \;
\left. \left[  x g(x,\mu^2) \right]^2\right\vert_{x =Xu} \; .
\label{lowest}
\eeq
Next, consider the limit
\beq
1 \ll \Delta \ll \log\left({\sqrt{s}\over M}\right) \; .
\eeq
The first term on the right-hand side of Eq.~(\ref{lowest}) tends to
a finite value of
\beq
{ \overline{\sum}\;| {\cal M}(gg\to gg)|^2  \over
  256  \pi \, \cosh^4(\half\Delta) } \ \to \ \half \pi C_A^2   \alpha_s^2 \; .
\label{amplim}
\eeq
The integral over the parton distributions is dominated by the contribution
 from the lower limit, i.e.
\beq
 \int_1^{X^{-2}} \; {\d u^2\over u^4} \;
\left. \left[  x g(x,\mu^2) \right]^2\right\vert_{x =Xu} \ \to \
\left[  X g(X,\mu^2) \right]^2
\label{partonlim}
\eeq
Combining the results from Eqs.~(\ref{amplim},\ref{partonlim}) gives
\beq
\left. {\d\sigma\over \d y_1 \d y_2 }\right\vert_{y_1 = -y_2 =
\half\Delta} \simeq  \half \pi C_A^2   \alpha_s^2 \; {1\over M^2}
\; \left[  X g(X,\mu^2) \right]^2 \;
\label{asymp222}
\eeq
in agreement with the leading-order contribution in Eq.~(\ref{basica}).

The next step is to investigate quantitatively how rapidly
the asymptotic result is attained in practice. We focus first on the matrix
element part, Eq.~(\ref{amplim}). Figure~1 shows the ratio of the left-hand
side (the exact result) to the right-hand side (the asymptotic result), as a
function of the rapidity gap $\Delta$.
Evidently,  the two agree to better than $10\%$ for $\Delta > 3.3\, $.
This is not unexpected, since we are testing here the size of the
\lq power correction' terms of order $e^{-\Delta}$.
We can also test the effective subprocess approximation, by comparing
the $qg\to qg$ and $\qq\to\qq$ amplitudes,
scaled by $9/4$ and $(9/4)^2$ respectively,
to the asymptotic  $gg\to gg$ result. These ratios are shown
as the dashed ($qg$) and dash-dotted  ($\qq$) lines in Fig.~1.
The approach to the limiting form is very similar to the $gg$ amplitude,
indicating that in the large $\Delta$ region the effective subprocess
approximation  is valid.

We have already discussed how when $M$ and $\sqrt{s}$ are fixed
 the bulk of the $\Delta$ dependence comes from the parton distributions.
 This is illustrated in Fig.~2, where the cross section of Eq.~(\ref{lowest}),
  scaled by $M^2$, is shown as a function of $\Delta$. We have chosen
 $\sqrt{s}= 1.8\ \TeV$, and used the latest \lq MRS(H)' partons \cite{MRSH}
 with $\lmsb^{(4)} = 230\ \MeV$. The scales in the running coupling
 and in the parton distributions are both set equal  to $M$. The
 solid curves correspond to the exact $gg\to gg$ cross section,
  for the representative values $M=10\ \GeV$ and $M=30\ \GeV$.
Also shown (dashed lines) are the asymptotic cross sections defined
by Eq.~(\ref{asymp222}).  There is a broad range of $\Delta$ where
the shape of the exact cross sections is reasonably well approximated.
The normalization, however, is too high by a factor
of order two.\footnote{In the range $2 < \Delta < 6$ for $M= 10\ \GeV$
the ratio of the
exact to the approximate  cross section always lies in the range 0.45 to 0.55.}
This can be traced to the fact that Eq.~(\ref{partonlim}) is only valid when
both $X$ is small {\it and}  the function $xg(x,\mu)$ is slowly varying.
If $xg \sim x^{-\lambda}$ in the relevant $x$ region, then an error
of order $(1+\lambda)$ is made  in the normalization.

\subsection{Comparison with exact $O(\as^3)$ calculation for $\phi \neq 0$}
Expanding the exponential inside the integral  in Eq.~(\ref{coeffs}) allows
us to read off the  differential $\phi$-distribution at next-to-lowest
order, i.e. $O(\as^3)$:
\beq
M^2 \left. {\d \sigmahat\over \d\phi}\right|_{\rm NLO} =
\left( {\as C_A\over\pi}\right)^3
\; {\pi^3\over 2}\; \Delta \;   f(\phi) \; ,
\label{signlo}
\eeq
where
\beq
f(\phi) = {1\over 2\pi}\;\sum_{n = -\infty}^{+\infty} \;
e^{in\phi} \;c_n \; ,
\label{series}
\eeq
and the Fourier coefficients are
\beq
c_n  = {1\over 2\pi}\; \int_{-\infty}^{+\infty} \; {\d z\over z^2
+ \quarter } \; \chi_n(z) \;  = \; 2\; [\; \psi(1) -  \psi(1+\half |n|) \; ]\;
{}.
\label{coeffnlo}
\eeq
The coefficients can be calculated explicitly by contour integration in the
complex $z$ plane:
\beqn
c_{n} = c_{-n}\; ,\; & &\;  c_{n+2} = c_n - {4\over 2+n}\ \  (n\geq 0)\; ,
\nonumber \\
c_0 = 0\; ,\;  & &\;  c_1 = 4\, ( \log 2 - 1) \; .
\eeqn
The first point to note  is that since  $c_0 = 0$, the integral of $f(\phi)$
vanishes, i.e.
\beq
M^2\;  \left.\sigmahat\right|_{\rm NLO} = 0\; .
\eeq
To calculate the  $\phi$ distribution we have to
 substitute
the $c_n$ coefficients into Eq.~(\ref{series}) and sum the Fourier series. We
have found it easier, however, to start from the solution to the
BFKL equation in transverse momentum space, where for $\phi \neq 0$ (see
Appendix),
\beqn
f(\phi) & = & {M^2 \over 2\pi}\; \int_{M^2}^{\infty}\int_{M^2}^{\infty}\;
{ \d k_{T1}^2\; \d k_{T2}^2\over k_{T1}^2 k_{T2}^2 (k_{T1}^2  +
k_{T2}^2 -2 k_{T1}k_{T2} \cos\phi) }  \nonumber \\
&=& {1\over \pi}\; \left[ \log(2(1-\cos\phi)) \; +\; (\pi-\phi)
\cot\phi \right]\ \quad \mbox{for}\ 0 < \phi \leq \pi\; ,
\label{magic}
\eeqn
and $f(-\phi) = f(\phi)$.
The singular behaviour $f\sim \phi^{-1}$  as $\phi \to 0$, which
corresponds to  almost
back-to-back jets, arises from  soft emission of the third (gluon) jet
 \cite{BFKL}.
It is cancelled in the total cross section by a virtual gluon contribution
proportional to $\delta(\phi)$. This can be taken into account
by invoking
the standard `plus prescription', i.e. $f(\phi) \to [f(\phi)]_+$ where
\beq
\int_{-\pi}^{\pi}\;  \d\phi\;  g(\phi)\;  [f(\phi)]_+
= \int_{-\pi}^{\pi}\;  \d\phi\;  (g(\phi) - g(0)) \;  f(\phi) \; .
\eeq

It is important to note that the distribution we have derived is
a leading logarithm result, in that it corresponds to retaining only
the leading $\as\Delta $ contribution and ignoring
corrections of order $\as$,  $\as e^{-\Delta}$ etc. To study the validity
of this approximation, we can  compare the result  in Eq.~(\ref{magic}) with
an exact calculation based
on the complete $gg\to ggg$ matrix element. The analogue of the $2\to 2$
cross section Eq.~(\ref{twototwo}) is
\beqn
\left. { \d\sigma\over \d y_1 \d y_2 \d\phi  }\right\vert_{y_1 = -y_2 =
\half\Delta} &= & {1\over 512\pi^4}\;
 \int_{M^2} \d k_{T1}^2  \;
 \int_{M^2}  \d k_{T2}^2  \;
 \int_{-\half\Delta}^{\half\Delta}  \d y_3 \nonumber \\
& & \times  \; [ x_1 g(x_1,\mu^2)\; x_2 g(x_2,\mu^2) ] \nonumber \\
 & & \times \;  \hat{s}^{-2}\;
  \overline{\sum}\;| {\cal M}(gg\to ggg)|^2
\label{twotothree}
\eeqn
where
\beqn
k_{T3}^2 & = & k_{T1}^2 +k_{T2}^2 - 2 k_{T1}k_{T2} \cos\phi \nonumber \\
x_1 & = & (k_{T1} e^{\half\Delta} +
 k_{T2} e^{-\half\Delta}  + k_{T3} e^{y_3}   )/\sqrt{s}   \nonumber \\
x_2 & = & (k_{T1} e^{-\half\Delta} +
 k_{T2} e^{\half\Delta}  + k_{T3} e^{-y_3}   )/\sqrt{s}   \nonumber \\
\hat{s}&=& k_{T1}^2 + k_{T2}^2 +k_{T3}^2 + 2k_{T1}k_{T2}\cosh\Delta\nonumber \\
& & + 2k_{T1}k_{T3}\cosh(\half\Delta-y_3)
+ 2k_{T2}k_{T3}\cosh(\half\Delta+y_3) \; . \nonumber \\
\eeqn
According to the BFKL analysis, this cross section should have the
asymptotic limit
\beq
\left. { \d\sigma\over \d y_1 \d y_2\d\phi }\right\vert_{y_1 = -y_2 =
\half\Delta} = \;
\left( {\as C_A\over\pi}\right)^3
\; {\pi^3\over 2 M^2 }\; \Delta \;   f(\phi)
 \; [ X g(X,\mu^2)]^2 \; .
\label{twotothreeas}
\eeq
We can see how this behaviour arises: the matrix element in (\ref{twotothree})
is dominated by configurations where the third gluon jet is produced centrally,
i.e. $|y_3| \ll \half\Delta$. With the matrix element approximated by its
value at $y_3 = 0$, the $y_3$ integral gives the overall factor of $\Delta$,
and
the remaining $k_{Ti}$ integrals give the function $f(\phi)$, as in
Eq.~(\ref{magic}). The parton distributions are again dominated by their
values at $x_1 = x_2 = X = 2M\cosh(\half\Delta)/\sqrt{s}$, as for the
leading order cross section.

We can study the approach to the asymptotic result in two stages. First,
at the subprocess level, we can compare the cross section
\beq
\left. { \d\hat\sigma\over \d y_1 \d y_2 \d\phi}\right\vert_{y_1 = -y_2 =
\half\Delta} =  {1\over 512\pi^4}\;
 \int_{M^2}^{\infty} \d k_{T1}^2  \;
  \d k_{T2}^2  \;
 \int_{-\half\Delta}^{\half\Delta}  \d y_3  \;  \hat{s}^{-2}\;
  \overline{\sum}\;| {\cal M}(gg\to ggg)|^2
\label{twotothreesub}
\eeq
with the asymptotic form
\beq
\left. { \d\hat\sigma\over \d y_1 \d y_2 \d\phi}\right\vert_{y_1 = -y_2 =
\half\Delta} = \;
\left( {\as C_A\over\pi}\right)^3
\; {\pi^3\over 2 M^2 }\; \Delta \;   f(\phi)\; .
\label{assub}
\eeq
Figure~3 compares the $\phi$ distribution calculated from
Eq.~(\ref{twotothreesub})
with the function $f(\phi)$, for $\Delta = 4, 8, 12$.
 The exact calculation has
 been scaled by the same factors multiplying $f(\phi)$
 on the right-hand side of
Eq.~(\ref{assub}), so that the exact and approximate distributions
 should
coincide in the limit $\Delta \to \infty$. The results confirm the
approach to the asymptotic distribution. At small $\phi$, the
asymptotic behaviour is already a good approximation for $\Delta
=4$, while the convergence is slower at large $\phi$. This is
presumably because at large $\phi$ the third gluon can have significant
transverse momentum and energy, thus invalidating the approximations
under which the BFKL equation is derived and leading to sizeable
sub-asymptotic corrections.

Figure~4 makes the same comparison at the cross-section level,
for $p \bar p$ collisions
with $\sqrt{s} = 1.8\ \TeV$, $M= 10\ \GeV$ and $\Delta = 4,6,8$.
The curves correspond
to the exact and asymptotic cross sections of Eqs.~(\ref{twotothree})
and (\ref{twotothreeas}) respectively.
The constraints $x_1, x_2 \leq 1$ now give
upper limits on the transverse momentum integrals. As for the leading-order
case, the {\it normalization} is overestimated by the asymptotic
form, but evidently the {\it shape} of the $\phi$ distribution
is reasonably well approximated even at moderate $\Delta$.
The rapid change in the distribution with increasing $\Delta$ supports our
 assertion that the azimuthal distribution of the
jets should be a more reliable indicator of BFKL behaviour than the
overall $\phi$-integrated cross section.

\subsection{All-orders $\phi$ distribution}
Through next-to-lowest order, then,  we have
\beqn
M^2  {\d \sigmahat(E,M)\over \d\phi} &=
&\left( {\as C_A\over\pi}\right)^2
\; {\pi^3\over 2} \;   F(\phi, \Delta)  \nonumber \\
 F(\phi, \Delta) & = &  \delta(\phi) \; +\;
 \left( {\as C_A\over\pi}\right)
 \;  \Delta \; [f(\phi)]_+ \;+ \; \ldots  \; ,
\eeqn
where the $\ldots$ represent terms $O((\as\Delta)^n)$ with $n\geq 2$.
Formally, the first few terms in this series will be a good approximation to
the
all-orders distribution provided $\Delta \gg 1$ and $\as\Delta \ll 1$.
As the second of these inequalities is relaxed, higher order terms
become more and more important. The inclusion of all terms of the form
$(\as\Delta)^n$, via Eqs.~(\ref{fourier},\ref{coeffs}),
requires a numerical calculation and
will be discussed below. First, we consider the limit
$\as\Delta \gg 1$, where an analytic approximation can again be
obtained.

To calculate the distribution in the asymptotic
limit $\as\Delta \to \infty$, we return
 to Eq.~(\ref{coeffs}) and use a saddle-point method \cite{BFKL} to evaluate
the
Fourier coefficients in the large
$t = \as C_A \Delta / \pi$ limit. We  expand the $\chi_n(z)$
about the saddle point at $z=0$,
\beq
\chi_n(z) = a_n - b_n z^2 + \ldots \; ,
\eeq
where
\beqn
a_0 = 2\log 2, \ a_1 = 0, & \quad & a_{n+2} = a_n - {2\over 1+n}, \ \
(n\geq 0) \nonumber \\
b_0 = 7\zeta(3), \ b_1 = \zeta(3), & \quad & b_{n+2} = b_n - {8\over
(1+n)^3} , \ \ (n\geq 0)
\eeqn
 from which the asymptotic $t\to\infty$ behaviour follows,
\beq
C_n(t)\; \sim \; {1\over \sqrt{\half \pi b_n t }} \; e^{2 a_n t}\; .
\label{asymp}
\eeq
Figure~5 shows the first seven $C_n$ coefficients as  functions of $t$.
Note that because $a_n < 0$ for $n \geq 1$, all but the $C_0$
coefficient tend to zero as $t\to \infty$. The asymptotic $\phi$
distribution is then obtained by substitution in Eq.~(\ref{fourier}),
\beqn
F(\phi, \Delta) &\sim & {1\over 2\pi}\; \left[\;
  {1\over \sqrt{\half \pi 7 \zeta(3) t }} \; e^{4 \log2\; t}\;
    + \;
  {2 \cos\phi \over \sqrt{\half \pi \zeta(3) t }} \; \right. \nonumber \\
 & & \; \left.  + \;
{2 \cos 2\phi \over \sqrt{\half \pi (7\zeta(3)-8) t }}\; e^{4 (\log2 -1)\; t}
\;
   + \; \ldots\;
\right] \; .
\label{fourieras}
\eeqn
Asymptotically, then, the $\phi$ distribution becomes {\it flat}.
The emission of an infinite number of soft gluons has completely
smeared out  the
back-to-back correlation exhibited by the lowest contributions to
the perturbation series.
Note that we have also reproduced the  asymptotic  result for the
$\phi$-integrated cross section \cite{BFKL,MUENAV}:
\beq
M^2 \sigmahat(\as,M^2, \Delta) = \left( {\as C_A\over\pi}\right)^2 \;
 {\pi^2\over 4} \; C_0(t) \longrightarrow
\left( {\as C_A\over\pi}\right)^2 \;  {\pi^2\over 4} \;
{1\over \sqrt{\half \pi 7 \zeta(3) t }} \; e^{4 \log2\; t}\; .
\label{sigas}
\eeq
For large $\Delta$, therefore,  we obtain the result given in
Eq.~(\ref{lambda})
with
\beq
\lambda  = {\alpha_s \over \pi} \, 4\, C_A \, \log2     = 0.5 \; ,
\eeq
for $\alpha_s =  0.19$.
The $\phi$-integrated cross section
was studied in some detail in Ref.~\cite{MUENAV}, where
an analytic approximation valid for $t \lapproxeq 1$  was derived.
Figure~6 shows the function $C_0(t)$, (i) computed exactly using
Eq.~(\ref{coeffs}) (solid line), (ii) according to the analytic approximation
of Ref.~\cite{MUENAV} (dotted line), and (iii) in the asymptotic
limit, Eq.~(\ref{sigas}) \cite{BFKL} (dashed line).

We have  so far obtained analytic approximations for the
small $t$ and large $t$ behaviours of the differential $\phi$
distribution. The distribution at arbitrary $t$ requires a numerical
calculation of the sum and integral in Eqs.~(\ref{fourier},\ref{coeffs}).
Thus, Fig.~7 shows the function $F(\phi, \Delta)$ defined in
Eq.~(\ref{fourier})
for $t = \as C_A \Delta /\pi = 0.25, 0.5, 1.0, 1.5$. We see very clearly
the transition from a sharply peaked distribution at small $t$ --- recall
that $ F = \delta(\phi) $ at $t=0$ --- to a larger, flatter distribution
as $t$ increases. As the rapidity gap widens, the emission of
more and more soft  gluons  uniformly
`fills in' the distribution at large $\phi$.

\section{Predictions for $p \bar p$ collisions at $1.8\ \TeV$}

The most direct test of the BFKL perturbative pomeron behaviour is the
rise in the subprocess cross section with
increasing rapidity gap $\Delta$, i.e.  $M^2\sigmahat \sim
\exp(\lambda\Delta)$.
However, as discussed in the Introduction, one cannot yet  regard this as
a precision prediction of the theory. One particular issue concerns
the inclusion of a running coupling in the BFKL analysis, i.e. $\alpha_s
\to \alpha_s(k_T^2)$. This prevents the integrals being extended down
to $k_T = 0$, thereby inducing a weak dependence on an infra-red cut-off
 parameter, see for example Ref.~\cite{AKMS}.
 Increasing  this cut-off reduces the phase space for the soft gluon emission
 and weakens the growth in the cross section with $\Delta$. In addition, the
 subleading logarithmic corrections to the BFKL result
  are not yet known.
As we have stressed, at fixed hadron collider energy the BFKL behaviour
is anyway  masked by additional dependence on $\Delta$  coming from the
parton distributions.
To investigate this latter effect quantitatively,
we show in Fig.~8 the cross section  of Eq.~(\ref{basic})
 at $\sqrt{s} = 1.8\ \TeV$
as a function of $\Delta$ for two choices of minimum jet transverse momentum,
$M =10\ \GeV$ and $M= 30\ \GeV$.
The dashed curves correspond to the leading order
 contribution to $\sigmahat$, i.e.
the first term on the right-hand side in Eq.~(\ref{defs}),
while the solid curves are the all-orders BFKL result,
corresponding to $\sigmahat$ given in Eq.~(\ref{sigas}).
The parton distributions are the latest MRS(H) set \cite{MRSH},
with $\lmsb^{(4)} =230\ \MeV$,
 which are consistent with both the recent HERA  and the fixed-target
  $F_2$ measurements.
Evidently, the shapes of the lowest-order and all-orders distributions
are quite different.
Notice that the $x$-dependence of the parton distributions more than
compensates  the BFKL rise in the subprocess cross section, so that the
net effect is a cross section which  {\it decreases}  as a function of
$\Delta$.
 However, we should recall from Section~\ref{locomparison}
that the lowest-order approximation is only a good representation
of the {\it shape} of the exact lowest order $2\to 2$ cross section for large
$\Delta\gapproxeq 4$.
 The signature for BFKL behaviour is therefore a slower fall-off
of the cross section with increasing $\Delta$ than predicted by the leading
order (exact or approximate) cross section. The size of the effect can be
gauged
 from Fig.~8. Whether the difference is detectable in practice depends
on the precision with which jets at large rapidity
can be reconstructed and measured experimentally.

We turn next to the distribution in the
 azimuthal angle difference of the two jets.
Figures 9 and 10 show the  $\phi = \pi - \phi_{jj}$
 distributions, at fixed $\Delta$, for $\sqrt{s}= 1.8\ \TeV$ with (a)
 $M = 10\ \GeV$ and  (b) $M = 30\ \GeV$.
In Fig.~9 the cross section itself is shown, while in Fig.~10
the distributions are normalized to have unit area for each value of $\Delta$.
The trend is that as $\Delta$ increases, the cross sections get smaller
and the distributions become flatter in $\phi$. The higher
the transverse momentum cutoff $M$, the faster the decrease with $\Delta$ and
the slower
the approach to the flat distribution.
For $M = 10\ \GeV$, values of $\Delta$ up to about 8 appear to be accessible,
at least in principle. For this rapidity gap, the  $\phi$ distribution is
almost
flat. A simpler representation of the flattening behaviour is provided by the
{\it average } of $\cos\phi$, which is proportional to  the $C_1$
 coefficient of Eq.~(\ref{coeffs}), i.e.
\beq
\langle \cos\phi \rangle \; = \; { C_1(t) \over C_0(t) } \; ,
\qquad t = {\alpha_s(M^2) C_A \Delta \over \pi}
\eeq
Figure 11 shows this average as a function of $\Delta$, for $M = 10\ \GeV$
and $M = 30\ \GeV$. The approach to flatness ($\langle \cos\phi \rangle \to 0$)
is slower for the higher cut-off because of the smaller coupling constant.

\section{Conclusions}

In this study we have focussed on two important aspects of BFKL perturbative
pomeron behaviour. The first  is the rise in the cross section
as the rapidity gap between two moderate $p_T$ jets increases, as first
discussed
in Ref.~\cite{MUENAV}. The difficulty with this signature is that in order
to circumvent the additional dependence  on rapidity induced by the parton
distributions, it is necessary to scale up the collider energy  as the rapidity
gap increases. At fixed collider energy, the dominant effect at large
rapidity gap is the suppression of the cross section by the fall-off
in the parton distributions as $x\to 1$. After allowing for the effects
of jet reconstruction and measurement in the detector,
it is not clear whether  the relatively small  effects of the BFKL
behaviour can be observed. This is not, however, as severe a
 problem for the second important
feature of the BFKL pomeron --- the weakening of the correlation in the
azimuthal angle of the two jets. The distribution in the
 azimuthal angle difference  $\phi$  changes
  from being back-to-back for central jet pairs with a small rapidity
difference,
  to an asympototically flat distribution as the jets separate in rapidity.

We have presented predictions  for the $\phi$ distribution at different
rapidity gaps $\Delta$ in $p \bar p$ collisions at $\sqrt{s} = 1.8\ \TeV$,
based on the fixed-coupling solutions of the BFKL equation.  However,
there are very likely non-negligible sub-leading corrections to this
behaviour. We have investigated some of these by comparing the
leading-logarithm
predictions with those based on the exact low-order matrix elements.
{}From this comparison, it seems that the leading behaviour may already
 be dominant for rapidity gaps as small as 4.

 Of course, we have {\it not} included such important effects as
 smearing of the jet energies and angles by the jet algorithms used
in the actual experiments.  A more precise analysis would
require a correspondong smearing of the $\phi$ distribution, which would
inevitably weaken
the correlations implied by perturbation theory alone. The size
of this smearing could be estimated, for example, by comparing
the actual $\phi$ distribution of two central jets  --- which should be
 well-described by lowest order perturbation theory ---  with the naive
 $\delta(\phi)$ expectation. If the smearing was parametrizable by, say,
 a gaussian distribution, this could be folded in to the Fourier
 coefficients in the perturbative predictions. We have not performed
such an analysis here, since the form of the smearing is presumably
detector and jet algorithm dependent.

In summary, therefore, it would be interesting to measure the
azimuthal angle distribution of the two-jet inclusive cross section
as a function of the jet rapidity gap, to see if the data are at least
qualitatively in line with the flattening
of the distribution predicted by the BFKL equation.  In this study
we have only considered the case of $p \bar p$ collisions at $1.8\ \TeV$.
However, basically the same behaviour should also be manifest
in  any high-energy collider with quarks and gluons in the initial state.
In particular, photoproduction of jet pairs at HERA could also be a useful
place to look for evidence of the BFKL behaviour, and, of course, high-energy
proton-proton colliders such as the LHC will allow a much larger range of
rapidities to be covered.

\vskip 1cm

\setcounter{section}{1}
\setcounter{equation}{0}
\renewcommand{\thesection}{\Alph{section}}
\renewcommand{\theequation}{\Alph{section}\arabic{equation}}
\section*{Appendix: Solution of the BFKL equation}

In this Appendix, we present a brief derivation of the
solution of the BFKL equation for the two-jet inclusive cross section.
More details can be found in Refs.~\cite{BFKL,MUENAV}.

We start from the subprocess cross section $gg\to gg +X$, where the two
final state gluons are produced with transverse momenta $k_{Ti}$ at large
rapidity separation $\Delta$, and $X$ represents additional soft gluons.
The differential cross section can be written, following the notation
of \cite{MUENAV},
\beq
{\d\hat\sigma\over \d^2k_{T1}\d^2k_{T2} } = {\as^2C_A^2 }
\; { f(k_{T1},k_{T2}, \Delta) \over k_{T1}^2 k_{T2}^2 }\;  .
\eeq
The Laplace transform of the function $f$,
\beq
\tf(k_{T1},k_{T2}, \omega) = \int_0^\infty \d\Delta
\; e^{-\omega\Delta}\; f(k_{T1},k_{T2}, \Delta) \; ,
\label{laplace}
\eeq
satisfies the BFKL equation \cite{BFKL}
\beqn
\omega \tf(k_{T1},k_{T2}, \omega) &=&
\delta(k_{T1}^2 -k_{T2}^2)\delta(\phi_1-\phi_2)
+\left({\as C_A \over \pi^2} \right)   \nonumber \\
&&\times \int{ \d^2k_T\over (k_{T1}-k_T)^2}
\left[ \tf(k_{T},k_{T2}, \omega) - {k_{T1}^2\tf(k_{T1},k_{T2}, \omega)
\over k_T^2 + (k_{T1}-k_T)^2 }\right] .
\label{bfkleqn}
\eeqn
This equation can be solved in closed form by introducing the Fourier
transform of $\tf$ with respect to $\phi_1 - \phi_2$ and $\log(k_{T1}^2/
k_{T2}^2)$:
\beq
 \tf(k_{T1},k_{T2}, \omega) = {1\over 2\pi}\sum_n e^{in(\phi_1-\phi_2)}
 \ {1\over 2\pi}\int_{-\infty}^{\infty} \d z\;  e^{-iz\log(k_{T1}^2/
k_{T2}^2)}\ \tf_n(z,\omega).
\eeq
Substituting this into Eq.~(\ref{bfkleqn}) gives
\beq
\omega \tf_n(z,\omega) = \left( k_{T1}^2 k_{T2}^2\right)^{-\half}
\; + \;\omega_0(n,z)\; \tf_n(z,\omega) \; ,
\eeq
where
\beq
\omega_0(n,z)\ =  \left({\as C_A \over \pi} \right) \;
2 \chi_n(z) \; ,
\eeq
with the function $\chi_n$ given in Eq.~(\ref{coeffs}). Performing
the inverse transform of Eq.~(\ref{laplace}) then gives
\beq
 f(k_{T1},k_{T2}, \Delta)  =  {1\over 2\pi}\sum_n e^{in\phi}
 \ {1\over 2\pi}\int_{-\infty}^{\infty} \d z  \;
(k_{T1}^2)^{-\half -iz} (k_{T2}^2)^{-\half +iz}
 \;  e^{2 t \chi_n(z)}  \; ,
\eeq
where $\phi = \phi_1 - \phi_2$ and $t=\as C_A \Delta /\pi$.
The final step is to integrate the transverse momenta
over the range $M^2 < k_{Ti}^2 < \infty $,
\beqn
\sigmahat(\as, M^2, \Delta) &=& {\pi\over 2} \int_{M^2}^\infty \d k_{T1}^2
\int_{M^2}^\infty \d k_{T2}^2 \int_{-\pi}^\pi \d\phi \
{\d\sigmahat\over \d^2k_{T1}\d^2k_{T2} }  \nonumber \\
&=& {\as^2C_A^2 \pi\over 2 M^2 }\; \int_{-\pi}^\pi \d\phi \
{1\over 2\pi}\sum_n e^{in\phi}
 \ {1\over 2\pi}\int_{-\infty}^{\infty}
{\d z\over z^2
+ \quarter } \; e^{2 t \chi_n(z)}\; ,
\eeqn
 which gives the
subprocess cross section  of  Eqs.~(\ref{sig},\ref{fourier},\ref{coeffs}).

\vskip 1cm

\section*{Figure Captions}

\begin{itemize}

\item [{[1]}]
Ratio of the exact $2\to 2$ subprocess cross sections defined in the text
(Eq.~(\ref{amplim}))
to the asymptotic $gg\to gg$  scaling form given in Eq.~(\ref{lo}),
  as a function
 of  the rapidity gap $\Delta$. The curves are (i) $gg\to gg$ (solid line),
 (ii) $q g \to q g$ multiplied by $9/4$ (dashed line), and
 (iii) $q \bar q\to q \bar q$ multiplied by $(9/4)^2$ (dot-dashed line).

\item [{[2]}]
 The lowest order $2\to 2$ cross section of Eq.~(\ref{lowest})
as a function of the rapidity gap $\Delta$, at
 $\sqrt{s}= 1.8\ \TeV$.
The  MRS(H) parton distributions \cite{MRSH}
 with $\lmsb^{(4)} = 230\ \MeV$ are used.
The   solid curves correspond to the exact $gg\to gg$ cross section
with minimum jet transverse momenta  $M=10\ \GeV$ and $M=30\ \GeV$.
Also shown (dashed lines) are the asymptotic cross sections defined
by Eq.~(\ref{asymp222}).

\item [{[3]}]
The  asymptotic azimuthal angle distribution $f(\phi)$
of Eq.~(\protect{\ref{magic}}), compared to the distribution
 calculated from the exact $gg\to ggg$ matrix
element for $\Delta = 4, 8, 12$ (dashed lines).

\item [{[4]}]
The dependence of the differential cross section at $O(\as^3)$ on
the azimuthal angle difference of the two jets with rapidity
gap $\Delta = 4,6,8$, in $p \bar p$ collisions
at $\sqrt{s} = 1.8\ \TeV$ with $M = 10\ \GeV$.
The  MRS(H) parton distributions \cite{MRSH}
 with $\lmsb^{(4)} = 230\ \MeV$ are used.
The solid curves correspond to the exact cross section
(Eq.~(\ref{twotothree})),
and the dashed curves to the asymptotic approximation of
Eq.~(\ref{twotothreeas}).

\item [{[5]}] The Fourier coefficients $C_n$  ($n\leq 6$)
defined in Eq.~(\ref{coeffs}) as  functions of
$t = \as C_A\Delta/ \pi $.

\item [{[6]}]
The Fourier coefficient  $C_0(t)$, which gives the $\phi$-integrated
cross section, (i) computed exactly using
Eq.~(\protect{\ref{coeffs}}) (solid line), (ii)
according to the analytic approximation
of Ref.~\cite{MUENAV} (dotted line), and (iii) in the asymptotic
limit, Eq.~(\ref{sigas}) (dashed line).

\item [{[7]}]
The azimuthal angular distribution function $F(\phi, \Delta)$,
 defined in Eq.~(\protect{\ref{fourier}}),
for $t = \as C_A \Delta /\pi = 0.25, 0.5, 1.0, 1.5$.

\item [{[8]}]
The cross section  $\d\sigma/\d y_1 \d y_2(y_1 = -y_2 = \half\Delta)$
of Eq.~(\ref{basic})  at $\sqrt{s} = 1.8\ \TeV$
as a function of $\Delta$, for two choices of minimum jet transverse momentum,
$M =10\ \GeV$ and $M= 30\ \GeV$.
The dashed curves correspond to the leading order
 contribution to $\sigmahat$, i.e.
the first term on the right-hand side in Eq.~(\ref{defs}),
and the solid curves are the all-orders BFKL result,
corresponding to $\sigmahat$ given in Eq.~(\ref{sigas}).
The  MRS(H) parton distributions \cite{MRSH}
 with $\lmsb^{(4)} = 230\ \MeV$ are used.

\item [{[9]}]
The  $\phi = \pi - \phi_{jj}$
 distributions, at fixed $\Delta$, for $\sqrt{s}= 1.8\ \TeV$ with (a)
 $M = 10\ \GeV$ and  (b) $M = 30\ \GeV$.
The  MRS(H) parton distributions \cite{MRSH}
 with $\lmsb^{(4)} = 230\ \MeV$ are used.

\item [{[10]}]
As for Fig.~9, but with the distributions normalized to unit area
at each $\Delta$.

\item [{[11]}]
The average azimuthal angle difference
$\langle \cos\phi \rangle  = -\langle \cos\phi_{jj} \rangle $
as a function of $\Delta$, for $M = 10\ \GeV$
and $M = 30\ \GeV$.

\end{itemize}

\end{document}